\documentclass[12pt,preprint]{aastex}
\usepackage{epsfig}

\def\part{\partial}

\def\bfj{{\bf j}}
\def\bfb{{\bf b}}

\def\beq{\begin{equation}}
\def\ee{\end{equation}}

\def\curl{\nabla \times}

\def\bfB{{\bf B}}

\def\bV{\overline V}

\def\bF{{\overline F}}
\def\bbV{{\overline {\bf V}}}
\def\bfv{{\bf v}}

\def\beqn{\begin{eqnarray}}
\def\eeqn{\end{eqnarray}}

\def\.{\mathaccent 95}
\def\beq{\begin{equation}}
\def\ee{\end{equation}}

\def\bB{\overline B}
\def\ts{\times}
\def\lb{\langle}
\def\rb{\rangle}
\def\curl{\nabla {\ts}}

\begin{document}
\title{Empirical relation between  
angular momentum transport and thermal-to-magnetic pressure ratio  
in shearing box simulations}

\author{E.G. Blackman$^1$ R.F. Penna$^{1}$, P. Varni\`ere$^{1,2}$}
\affil{1. Dept. of Physics and Astronomy, University of Rochester,
Rochester NY 14627, USA; 2.LAOG, Universit\'e J. Fourier UMR 5571, France}

\begin{abstract}


By combining data from different published 3-D simulations 
of Keplerian shearing boxes unstable to the magnetorotational instability (MRI), 
we highlight  tight anti-correlations  
between  the total effective inferred angular momentum transport  
parameter, $\alpha_{tot}$, 
its separate Maxwell and Reynolds contributions $\alpha_{mag}$ and $\alpha_{kin}$,  and the  kinetic to magnetic pressure ratio
$\beta$, defined with the initial or saturated (when available) 
thermal pressure.
 Plots of  $Log (\alpha_{kin}), Log (\alpha_{mag})$, and $Log (\alpha_{tot})$ 
vs $Log (\beta)$  are well fit by  straight lines 
even as $\alpha_{kin}$, $\alpha_{mag}$,and $\alpha_{tot}$ vary by four
orders of magnitude over the simulations included.
The ratio $\alpha_{kin}/\alpha_{mag}$ and the product $\alpha_{tot}\beta$ are 
quite constant and largely independent of the presence or absence of weak 
mean fields, the choice of initial and boundary conditions, and the resolution.
In short, simulations have more strongly constrained the product $\alpha_{tot}\beta$ than   $\alpha_{tot}$ itself.

\end{abstract}

\noindent {\bf Key Words}: accretion disks--magnetic fields--magnetohydrodynamics (MHD)--instabilities

\section{Introduction}

Accretion disks are widely appreciated to be a source of
emission from gas or plasma orbiting central stars or compact objects
(c.f. Frank, King, Raine 1992).  
In order to explain the rapid variability and short lifetimes of 
accreting systems without unphysical mass densities, some enhanced
angular momentum transport beyond that which can 
be supplied by the microphysical transport coefficients is typically required.
For sufficiently ionized disks, the magneto-rotational instability (MRI)
offers a solution to this problem for sufficiently ionized disks 
(e.g. Balbus \& Hawley 1991; 1998).

The MRI feeds off of an initially weak magnetic field
and the turbulence induced by the ensuing instability amplifies
the fluctuating magnetic energy by line stretching. 
Sustained magnetic fields under the influence of a shear flow 
in a radially decreasing  angular velocity profile  
produce a negative magnetic (Maxwell) stress, 
which, in principle, produces the dominant  
positive outward angular momentum transport. 3-D Simulations 
(e.g. Hawley, Gammie, Balbus, 1995,1996;
Brandenburg et al. 1995; Stone et al. 1995)
have revealed that the nonlinear evolution of systems
unstable to the MRI  leads to a Maxwell stress
whose magnitude is larger than the negatively
signed Reynolds stress. The MRI sustains the 
turbulent Maxwell stress and thus the outward
angular momentum transport.

While the MRI has been numerically shown to provide an effective
turbulent magnetic stress, incorporating the saturated state of the MRI
into the framework of practical accretion disc modeling using, for example
the $\alpha_{tot}$ viscosity coefficient formalism 
(Shakura \& Sunyaev 1973), 
where $\alpha_{tot}$ is defined from the turbulent viscosity,
$\nu_T =\alpha_{tot} c_s H$
and $c_s$ and $H$ are the sound speed and disk scale height)
suffers from the non-universality of 
values of $\alpha_{tot}$  inferred from simulations. 
Depending on the boundary conditions, initial conditions, different
treatments of viscosity and resistivity, 
the presence or absence of stratification, and resolution (Pessah et al. 2007), the 
values inferred from simulation can vary by 4 orders of magnitude
(see Figs. 1 and 2 below).
However, 
the $\alpha_{tot}$ prescription provides a  practical mean field formalism that 
allows a straightforward  calculation of accretion disk spectra for
comparison to observations by parameterizing nonlinear correlations 
of turbulent fluctuations by a simple closure.
Developing an improved 
mean field theory that also incorporates the physics of the MRI, while still
being practical is an important target of recent and ongoing work (e.g.  
Ogilvie 2003; Pessah 2006ab).

Here  we emphasize that the ratio of the thermal to magnetic
pressure, $\beta$  is not generally an independent function of $\alpha_{tot}$, 
even though it is sometimes
assumed to be in phenomenological analytic disc models 
(e.g. Yuan et al. 2005).  In this paper 
we combine the published data extracted into  Tables 1-10
to determine the empirical correlation between the kinetic and magnetic
contributions to $\alpha_{tot}$. 

In Sec. 2 we derive the formalism that relates the kinetic and magnetic
parts of $\alpha_{tot}$ to $\beta$ for different adiabatic indices
and give a physical argument for an 
inverse relation between $\alpha_{tot}$ 
and $\beta$. We  do not 
present a  rigorous theory in the present work as
our main focus is empirical.  Toward this end, 
in Sec. 3  we plot the data points from published
simulations and infer the empirical values for the quantities 
defined in Sec. 2. The data reveal that the product $\alpha_{tot}\beta$  
is nearly constant. We conclude in Sec. 4.

\section{Maxwell and Reynolds Contributions to Transport
and Relations to $\beta$}

In the steady-state,  ignoring microphysical viscosity, 
the  mean azimuthal momentum equation is given by (e.g. 
Balbus \& Hawley 1998)
\begin{equation}
\nabla\cdot 
\left[r\rho v_\phi {\bf v}-
r {B_\phi\over 4\pi}\bfB_p
+r\left(P+{B_p^2\over 8\pi}\right){{\bf e}_\phi}\right]=0
\end{equation}
The quantity inside the brackets represents the flux
of angular momentum. Of particular interest is the $r\phi$
component of this flux, which, when greater than zero, represents the outward 
radial transport of angular momentum. It is given by 
\beq
F_{r\phi} =\left[r\rho v_\phi {v}_r-
r {B_\phi\over 4\pi}B_r\right].
\label{2}
\ee

Because  axisymmetric accretion disk equations formally represent 
mean field equations, we are interested in the averaged value
of $F_{r\phi}$. Toward obtaining this, we 
split the magnetic field and velocity into mean (indicated by 
an overbar) and fluctuating components (indicated by lower case).
As in Balbus \& Hawley (1998), we take the mean to represent a
height integration over all $z$, an 
average over all $\phi$ and an average over some fixed
range of $r$. For a quantity $Q={\overline Q}+q$ we have
$\lb q\rb =0$ and 
\beq
\lb Q \rb ={\overline Q} =
{\int Q\rho d\phi dr dz \over 2\pi \Sigma \Delta r},
\label{3}
\ee
where $\Sigma =\int_{-H}^H \rho dz$.
Assuming that $\rho={\overline \rho}$ (no fluctuating density),
applying (\ref{3})  to (\ref{2}) gives
\beq
 {\overline F}_{r\phi} = {\Sigma r\over 2H}
\left[
\bV_\phi \bV_r-
 \bV_{A,\phi} \bV_{A,r}
+ \lb v_\phi v_r\rb - \lb b_{\phi} b_{r}\rb
\right],
\label{4}
\ee
where $\bbV_A$ is the 
Alfv\'en velocity associated with the
mean field and $\bfb$ is the 
Alfv\'en velocity associated with the fluctuating field.
The first term on the right is an inward flux of angular momentum, 
since $\bV_r < 0$ and $\bV_\phi >0$ at the inner most radii for an
accretion disc.  The remaining terms must provide  the needed
outward transport of angular momentum if matter is to accrete.
The last two terms represent purely turbulent transport.
In what follows, 
we assume that the mean magnetic field of smaller magnitude than the
fluctuating field in saturation and that the dominant angular momentum
transport comes from the last two terms of (\ref{4}).(This is consistent
with all of the simulations we consider.)

The shearing box simulations of Table 1 
employ local Cartesian coordinates in the rotating frame.
In this shearing-sheet approximation, 
the mean velocity $\bV_y$ vanishes at the inner most radius $r_0$ 
of the shearing box, and  points in the $-\hat {\bf y}$ for $x=r-r_0>0$, 
decreasing outward in Keplerian
fashion such that   ${\bV}_y={(r-r_0)} r{\partial {\tilde\Omega}\over \partial r} 
\simeq  -{3\over 2} \Omega x$. Here  $\tilde \Omega$ is the local orbital
speed,  $\Omega$ is the orbital speed of the rotating frame, 
$x\equiv r-r_0$, and $x << r_0$.
In this context, we can combine the 
last 2 contributions of (\ref{4}) into a Cartesian 
stress tensor 
\beq
{\overline W}_{xy}\equiv 
 \lb v_y v_x\rb - \lb b_{y} b_{x}\rb.
\label{4a}
\ee
Using the Shakura-Sunayev prescription of $\nu_{T}\equiv \alpha_{tot} c_s H$ 
 of Sec. 1, 
the stress   for a Keplerian
flow in a shearing box that corresponds to an outward flux of angular
momentum is  
\beq
-\nu_{eff}\partial_x {\bV}_y\simeq {3\over 2}\Omega\alpha_{tot} c_s H .
\label{5a}
\ee
Setting this equal to (\ref{4a}) gives a closure for the stress tensor, 
so that 
\beq
\alpha_{tot} =
 {2{\overline W}_{xy}\over  3\Omega c_s H}=
{2f(\Gamma){\overline W}_{xy}\over 3c_s^2},
%
\label{6}
\ee
where $\Gamma$ is the polytropic index and 
$f(1)=
\sqrt{1\over 2}$ (isothermal) and  $f(5/3)=
\sqrt{1\over 3}$ (adiabatic).
 The  last relation in 
(\ref{6})
comes from solving the equation of hydrostatic equilibrium, 
namely 
\beq
{1\over {\overline \rho}} {\partial {\overline P}\over \partial z}\simeq -{GM\over R^2}{z\over R}.
\label{8a}
\ee
Over a density scale height, the solution gives
$\Omega H= {c_s\over f(\Gamma)}$, with midplane
sound speed $c_s$.

We now split (\ref{6}) into magnetic and kinetic terms such that 
$\alpha_{tot}=\alpha_{kin}+\alpha_{mag}$ where 
\beq
\alpha_{mag}
\equiv -
{2f(\Gamma)\lb b_{y} b_{x}\rb \over 3c_s^2}=
{C_{mag}(\Gamma,\beta)\over \beta},
\label{7}
\ee
and 
\beq
\alpha_{kin}\equiv {2f(\Gamma) \lb v_y v_x\rb \over 3c_s^2 }
\equiv
{C_{kin}(\Gamma,\beta)\over \beta},
\label{8}
\ee
where $C_{mag}(\Gamma,\beta)\equiv-{4f(\Gamma)\over 3\Gamma}
{\lb b_x b_y\rb\over \lb b^2\rb}$ and $C_{kin}(\Gamma,\beta)\equiv
{4f(\Gamma)\over 3\Gamma}{\lb v_x v_y\rb\over \lb b^2\rb}$
are to be inferred from the data.
We  also define
$C_{tot}(\Gamma,\beta)$ such that 
$\alpha_{tot}= C_{tot}/\beta$. The statistically 
determined $C_{tot}$ need not exactly equal the separately determined 
best fit values of $C_{mag}$ and $C_{kin}$.

We can provide  a crude physical argument 
which anticipates a strong 
anti-correlation between $\alpha_{tot}$ and $\beta$:
Note that 
$\nu_{eff}=\alpha_{tot}c_s H= v_T L,$ where $v_T$ and $L$ are a  turbulent
velocity and dominant energy containing fluctuation scale.
 In a turbulent flow, the ratio of  magnetic to kinetic turbulent energies
is typically of order unity in saturation 
(and actually slightly larger than unity for MRI simulations).
Crudely, if  $v_A \sim v_T$, then 
$\alpha_{tot}c_s H \sim v_A L$. But $L\sim v_T/\Omega \sim v_A/\Omega$, 
if the eddy turnover time scale is comparable to the orbit time.
The latter is a reasonable assumption since 
the growth rate for a MRI instability
that initiates turbulence is of order the rotation rate.
We therefore have $\alpha_{tot}c_s H\sim v_A^2/\Omega$
which implies $\alpha_{tot} \sim 
{2 f(\Gamma)\over \Gamma \beta}$,
using the relations below Eq. (\ref{8a}). 
The specific anisotropy due to Keplerian shear
likely implies a missing factor of order unity.


Note that 
$\alpha_{mag}$, $\alpha_{kin}$ 
$\alpha_{tot}$, and $\beta$ were defined above
 as a function of  the actual thermal pressure at the instant of measurement, not the initial thermal pressure of a system from which the system could evolve.  For the isothermal case, the distinction is not important because 
the pressure is a constant in time in the simulations, but for the adiabatic case the pressure evolves.  For most of the published simulations, the transport coefficients and $\beta$ are  expressed with respect to the initial pressure at the midplane $P_0\equiv P(t=0,z=0)$, where the subscript indicates both the initial time and 
the midplane $(z=0)$.  We distinguish this from the saturated 
thermal pressure at the midplane  $P(0)\equiv P(t>>0,z=0)$.
For Sano \& Stone (2002),  data for of the transport coefficients and $\beta$ were published. Subsequently, we will explicitly distinguish 
$P_0$ and $P(0)$ where necessary.
For the isothermal case we note that $P_0= P(0)\equiv P(z=0)$ for all time.

\section{The $\alpha(\beta)$ Relation from Published Numerical Simulation Data}

We have extracted data from published shearing box 
simulations to produce Tables 1-10. Note that as emphasized in the table
captions, the double brackets used therein indicate the combination of 
a spatial average (as employed  above) along with  
a time average at late tmes in the simulations.

To make the numerical
coefficients from the published data
 correspond with those of the previous section for 
in  $\alpha_{mag}$ and $\alpha_{kin}$ we 
have  multiplied the values from those references 
by ${2 f(\Gamma)\over 3\Gamma}$ in Tables 1-10.
Note also that the convention used in Brandenburg 1995 (see Brandenburg 1998)
is a factor of ${2\over 3f(1)}\sim {0.48}$ smaller 
than the convention used in  the other references.
Therefore, to match our numerical coefficients  we have multiplied the 
$\alpha$ from that reference by 
by ${2\over 3}{f(\Gamma)\over \Gamma}{3\over 2 f(1)}={f(\Gamma)\over f(1)\Gamma}$ to construct our Table 1 for $\alpha_{mag}$
and $\alpha_{kin}$. Also, the value of $E_{mag}$ in
Brandenburg et al. (1995) must be divided by a factor of $2$ when placed
into our definition of $\beta$ for the magnetic pressure.   

The simulations of Brandenburg  et al. (1995) also differ from the others in that
vertical field boundaries were used at the top and bottom
of the box. This allowed mean field amplification, in contrast 
to the periodic boundary conditions used in the
other simulations. 
However, since the 
mean field saturates at values relatively small compared to the random field, 
the effect of mean field growth on the total stress should be 
relatively small.

In Fig. 1abc we plot respectively,
 $Log\ \alpha_{kin}(P(0))$, 
$Log\ \alpha_{mag}(P(0))$, and $Log\ \alpha_{tot}(P(0))$ 
 vs. $Log\ \beta(P(0))$ for the $\Gamma=1$ isothermal simulations, and 
plot $Log\ \alpha_{kin}(P_0)$ vs.
$Log\ \alpha_{mag}(P(0))$ in Fig. 1d.  In Fig. 2 we show the analogous
plots for the  adiabatic $\Gamma = 5/3$ cases. There the
distinction between $P(0)$ and $P_0$ is necessary.  
In Fig 2 we include the Sano et al. (2002) data points using $P_0$,
whereas in Fig. 3 we give the analogous plots using $P$ rather than $P_0$, 
for the cases in which  $\alpha_{kin}(P)$, 
$\alpha_{mag}(P)$, and $\beta(P)$
values are available.
Although the use of $P$ vs. $P_0$ does not change products of the form
$\alpha\beta$, the range of values 
over which the $x$ and $y$ axes range can differ.  
From Fig. 3 however, it can be seen that 
the best fit curves still show a tight correlation and that the $x$ and $y$ axes range over more than an order of magnitude supporting the same general conclusions as the previous two Figs.


The best fit solid lines  are shown, with the equations
for these lines at the top of each panel in Figs.1, 2, and 3.
For each panel, the fit at the top
is of the form 
$Log (\alpha)= A Log (\beta) + D$, shown at the top
with $A$ and $D$ constants.
We can use these fits to extract best fit values of $C_{mag}(\Gamma,\beta)$ and 
$C_{kin}(\Gamma,\beta)$ defined in  Sec.2.
Counter clockwise from the top right panel of Fig.1 we then have
\beq
C_{mag}(1)= 0.13\beta^{-0.03},
\label{ck53}
\ee
\beq
C_{kin}(1)= 0.02\beta^{0.12} 
\label{cm53}
\ee
\beq
C_{tot}(1)= 0.11\beta^{-0.06},
\label{cm10}
\ee
and for Fig 2. we have
\beq
C_{mag}(5/3)= 0.10\beta(P_0)^{-0.03},
\label{ck530}
\ee
\beq
C_{kin}(5/3)= 0.02\beta(P_0)^{0.01} 
\label{cm5300}
\ee
\beq
C_{tot}(5/3)= 0.08\beta(P_0)^{-0.07},
\label{cm1}
\ee
 where we have added the explicit functional dependence on $P_0$.

From Fig. 3 we have  
\beq
C_{mag}(5/3)
= 0.11\beta^{-0.03},
\label{sanom}
\ee
\beq
C_{kin}(5/3)
= 0.01\beta^{0.20}
\ee
and
\beq
C_{tot}(5/3)
= 0.14\beta^{-0.23}.
\label{sanotot}
\ee

The results for Fig 3 are based on only 8 points, yet still
the weak dependence of 
the near constancy of $C_{mag}$ and $C_{tot}$ is revealed in (\ref{sanom})
and (\ref{sanotot}). More robust
are  (\ref{cm10}) and (\ref{cm1}) which highlight a tight correlation.
The results imply that  $C_{tot}(1)\sim C_{tot}(5/3)(P_0)\sim 0.1$ 
even as $\alpha_{tot}(P_0)$ varies by more than 4 orders of magnitude.

The bottom right panels in Figs. 1 and 2 also highlight strict correlations between
$\alpha_{kin}$ vs. $\alpha_{mag}$.
Fig 1d shows that the best
fit for the isothermal data is 
$\alpha_{kin}=0.22\alpha_{mag}^{0.95}$ and Fig. 2d shows that for the
adiabatic case,  $\alpha_{kin}(P_0)=0.28\alpha_{mag}(P_0)$.
The near constancy of 
$\alpha_{kin}/\alpha_{mag}$ was
also noted by (Pessah et al. 2006ab).


\section{Conclusion}

We have highlighted that 
data from published shearing box MRI simulations
(Tables 1-10) show tight anti-correlations of
$\beta$ with $\alpha_{mag}$, $\alpha_{kin}$, and their sum $\alpha_{tot}$.
In particular, using our definitions, the product $\alpha_{tot}\beta \sim 0.1$, 
even as $\alpha_{tot}$ varies by over 4 orders of magnitude between
simulations.   The data were taken  from simulations invoking different codes,
different vertical boundary conditions, different initial conditions,
the presence or absence of stratification,  the presence
or absence of initial mean fields, explicit vs. numerical viscosity,
different polytropic indices, and different resolutions.
The data reveal that simulations have much more strongly constrained the product $\alpha_{tot} \beta$ than any particular value of $\alpha_{tot}$.

The fact that no universal value of 
$\alpha_{tot}$ emerges 
implies that the boundary and initial conditions, as well as
 the resolution (Pessah et al. 2007) are influencing its value.
Pessah et al. (2007) show the magnitude of the magnetic 
stress correlates strongly with the box size. This is troubling 
as it implies no robust value of the transport coefficient presently emerges from  simulations. In contrast,  
the narrow range of values for the constant  product $\alpha_{tot}\beta$ 
then must likely emerge from features identical in all simulations, such as the 
Keplerian shear profile
This is consistent with  expectations from analysis of the linear regime
(Pessah et al. 2006a),   and more general nonlinear 
closures (e.g. Ogilvie 2003; Pessah et al. 2006b). More work is needed to
determine the constant $\alpha_{tot}\beta$ from first principles.
Alternatively, the constants $\alpha_{tot}\beta$ and 
$\alpha_{mag}/\alpha_{kin}$ can be employed as a constraints for closures.

Modelers appealing to the MRI and extracting guidance
from simulations should treat $\alpha_{tot}$ and $\beta$ as
dependent parameters. The value of  $\alpha_{tot}\beta $
extracted  from Fig. 2 emerges as much more robust universal constraint,  
than any specific value of $\alpha_{tot}$.
A relation between $\alpha_{tot}$ and $\beta$
has been incorporated  into some some disk models 
(e.g. Narayan et al. 1998) but not others (e.g. Yuan et al. 2005).

Note also that the data herein are for  thin disks. While similar principles
would apply for thick disks, the numerical constants could be different.

{\bf Acknowledgments}:
We thank A. Brandenburg, O. Gressel,  A. Johansen, M. Pessah, V. Pariev, D. Uzdensky for comments
and discussions.
We acknowledge support from NSF grants AST-0406799, AST
00-98442, AST-0406823, NASA grant ATP04-0000-0016, the Aspen Center for Physics,and KITP of UCSB, with support from NSF Grant PHY-9907949.  We acknowledge support from the Laboratory for Laser Energetics.

\eject

\begin{deluxetable}{lccc}
\tablewidth{0.0pt}
\tablecolumns{4}
\tablecaption{Brandenburg et al. 1995}
\tablehead{
  \colhead{Run} &
  \colhead{$\langle\langle 8 \pi P_0/B^2\rangle\rangle $} &
  \colhead{$(2 f(\Gamma)/3\Gamma)\langle\langle - B_x B_y / 4 \pi P_0\rangle\rangle$} &  
  \colhead{$(2 f(\Gamma)/3\Gamma)\langle\langle\rho v_x \delta v_y / P_0\rangle\rangle$}}
\startdata 
 A     &    154 & 0.000734847 & 0.000244949 \\
 B     &    143 & 0.000342929 & 9.79796E-05 \\
 C     &    40    & 0.002008582 & 0.000489898 \\
 D     &    143 & 0.000685857 & 0.000146969 \\
 E2    &    118 & 0.001126765 & 0.000146969 \\     
 AD    &    40    & 0.001518684 & 0.000342929 \\
\enddata
\tablecomments{Runs with adiabatic equations of state and vertical initial
  fields. The double brackets indicate that 
all data are time- and volume-averages at late times over
  the turbulent layer of the disk.  The magnetic field
  energy, Maxwell stress, and Reynolds stress are all normalized with
  respect to the initial gas pressure at the midplane, $P_0$. The run
  AD includes ambipolar diffusion.}
\end{deluxetable}

\begin{deluxetable}{lccc}
\tablewidth{0.0pt}
\tablecolumns{4}
\tablecaption{Fleming et al.  2000}
\tablehead{
  \colhead{Run} &
  \colhead{$\langle\langle 8 \pi P_0/B^2\rangle\rangle $} &
  \colhead{$(2 f(\Gamma)/3\Gamma)\langle\langle - B_x B_y / 4 \pi P_0\rangle\rangle$} &  
  \colhead{$(2 f(\Gamma)/3\Gamma)\langle\langle\rho v_x \delta v_y / P_0\rangle\rangle$}}
\startdata 
 BZ1 & 2.2222E+00 & 5.8659E-02 & 1.2009E-02 \\
 BZ2 & 3.6101E+00 & 3.9260E-02 & 9.2376E-03 \\
 BZ3 & 7.1429E+00 & 2.0092E-02 & 5.0807E-03 \\
 BZ4 & 1.6129E+01 & 9.6995E-03 & 2.7713E-03 \\
 ZN1 & 1.0000E+02 & 1.0392E-03 & 5.3116E-04 \\
 ZN2 & 3.3333E+02 & 2.3094E-04 & 2.3094E-04 \\
 ZN3 & 3.3333E+02 & 2.0785E-04 & 1.5935E-04 \\
 Y1 & 1.4184E+01 & 6.9282E-03 & 1.7782E-03 \\
 Y2 & 1.6393E+01 & 6.0044E-03 & 8.3138E-04 \\
 Y3 & 2.0408E+01 & 2.3556E-03 & 1.5935E-04 \\
\enddata
\tablecomments{Runs with adiabatic equations of state.  Vertical field
  runs are labeled by the prefix BZ, zero net $z$ runs
  are labeled by the prefix ZN, and toroidal field runs are
  labeled by the prefix Y.  The double brackets indicate that all data are time- and volume-averages at
  late times over the turbulent layer of the disk.  The
  magnetic field energy, Maxwell stress, and Reynolds stress are all
  normalized with respect to the initial gas pressure at the midplane,
  $P_0$. These runs include the effects of Ohmic resistivity.}
\end{deluxetable}

\begin{deluxetable}{lccc}
\tablewidth{0.0pt}
\tablecolumns{4}
\tablecaption{Hawley, Gammie, Balbus 1995}
\tablehead{
  \colhead{Run} &
  \colhead{$\langle\langle 8 \pi P_0/B^2\rangle\rangle $} &
  \colhead{$(2 f(\Gamma)/3\Gamma)\langle\langle - B_x B_y / 4 \pi P_0\rangle\rangle$} &  
  \colhead{$(2 f(\Gamma)/3\Gamma)\langle\langle\rho v_x \delta v_y / P_0\rangle\rangle$}}
\startdata 
 Z3  & 0.0294  & 2.678905249 & 0.743396207 \\
 Z4  & 1.9608  & 0.069974853 & 0.016165808 \\
 Z5  & 2.6385  & 0.034410076 & 0.013625466 \\
 Z7  & 2.5641  & 0.050575884 & 0.014780167 \\
 Z9  & 0.9346  & 0.121474497 & 0.027019993 \\
 Z12 & 1.0811  & 0.110851252 & 0.030022214 \\
 Z15 & 0.7686  & 0.154037052 & 0.032331615 \\
 Z17 & 3.9370  & 0.029329394 & 0.006928203 \\
 Z18 & 2.4038  & 0.061430069 & 0.027019993 \\
 Z19 & 2.1505  & 0.066741691 & 0.018475209 \\
 Z20 & 7.6336  & 0.014087347 & 0.003925982 \\
 Z21 & 6.5359  & 0.017089568 & 0.004618802 \\
 Z22 & 2.5974  & 0.049190243 & 0.010623245 \\
 Z23 & 27.0270 & 0.004618802 & 0.00092376  \\
 Z24 & 76.9231 & 0.001385641 & 0.00023094  \\
 Z25 & 30.3030 & 0.003233162 & 0.001154701 \\
 Y1  & 13.8889 & 0.007159143 & 0.002309401 \\
 Y2  & 30.3030 & 0.003002221 & 0.00092376  \\
 Y3  & 8.6207  & 0.011316065 & 0.003695042 \\
 Y6  & 23.8095 & 0.003002221 & 0.001154701 \\
 Y7  & 9.2593  & 0.008544784 & 0.002309401 \\
 Y8  & 1.6260  & 0.019398969 & 0.005080682 \\
 Y9  & 8.6207  & 0.009468544 & 0.002309401 \\
 Y10 & 62.5000 & 0.00069282  & 0.00023094  \\
 Y11 & 16.9492 & 0.005542563 & 0.001847521 \\
 Y12 & 50.0000 & 0.001847521 & 0.00069282  \\
 Y13 & 14.0845 & 0.006928203 & 0.002078461 \\
 Y15 & 8.5470  & 0.010854185 & 0.003233162 \\
 Y17 & 185.185 & 0.000508068 & 9.2376E-05  \\
 Y18 & 3.8610  & 0.022632131 & 0.006235383 \\
 YZ1 & 2.9674  & 0.019860849 & 0.010392305 \\
 YZ2 & 3.9370  & 0.026789052 & 0.006235383
\enddata
\tablecomments{Runs with adiabatic equations of state. Vertical field
  runs are labeled by the prefix Z and toroidal field
  runs are labeled by the prefix Y.  Combined toroidal and
  vertical field runs are labeled by the prefix YZ. The double brackets inidcate that all data are time-
  and volume-averages at late times over the turbulent layer of the
  disk.  The magnetic field energy, Maxwell stress, and
  Reynolds stress are all normalized with respect to the initial gas
  pressure at the midplane, $P_0$.}
\end{deluxetable}

\eject

\begin{deluxetable}{lccc}
\tablewidth{0.0pt}
\tablecolumns{4}
\tablecaption{Hawley, Gammie, and Balbus 1996}
\tablehead{
  \colhead{Run} &
  \colhead{$\langle\langle 8 \pi P_0/B^2\rangle\rangle$} &
  \colhead{$(2 f(\Gamma)/3\Gamma)\langle\langle - B_x B_y / 4 \pi P_0\rangle\rangle$} &  
  \colhead{$(2 f(\Gamma)/3\Gamma)\langle\langle\rho v_x \delta v_y / P_0\rangle\rangle$}}
\startdata 
 R1 & 34 & 0.002817469 & 0.000969948 \\
 R2 & 63 & 0.001524205 & 0.000623538 \\
 R3 & 32 & 0.003140785 & 0.000969948 \\
 R4 & 67 & 0.001408735 & 0.00057735  \\
 R6 & 16 & 0.006189195 & 0.002332495 \\
 R7 & 250& 0.000392598 & 0.000161658 \\
\enddata
\tablecomments{Runs with adiabatic equations of state and random initial $B$.
  The double brackets indicate that all data are time- and volume-averages at late times over the
  turbulent layer of the disk.  The magnetic field energy,
  Maxwell stress, and Reynolds stress are all normalized with respect
  to the initial gas pressure at the midplane, $P_0$.}
\end{deluxetable}

\begin{deluxetable}{lccc}
\tablewidth{0.0pt}
\tablecolumns{4}
\tablecaption{Sano \& Stone 2002}
\tablehead{
  \colhead{Run} &
  \colhead{$\langle\langle 8 \pi P_0/B^2 \rangle\rangle $} &
  \colhead{$\langle\langle {- B_x B_y \over  4 \pi P_0}
{2f(\Gamma)\over 3\Gamma}\rangle\rangle$} &  
  \colhead{$\langle\langle\rho v_x \delta v_y / P_0\rangle\rangle{2f(\Gamma)\over 3\Gamma}$}}
\startdata 
Z02 & 2.8094E+00 & 3.7412E-02 & 7.8058E-03 \\
Z03 & 4.2117E+00 & 2.4942E-02 & 5.0345E-03 \\
Z04 & 1.1940E+01 & 9.0990E-03 & 2.8637E-03 \\
S01 & 2.1511E+01 & 3.8798E-03 & 3.0022E-03 \\
S03 & 7.7320E+00 & 1.2009E-02 & 3.0715E-03 \\
S10 & 1.1988E+01 & 7.5286E-03 & 2.0346E-03 \\
Y02 & 1.2686E+01 & 8.1522E-03 & 1.9237E-03 \\
Y04 & 7.9931E+00 & 1.2494E-02 & 2.8868E-03 \\
\enddata
\tablecomments{Runs with adiabatic equations of state.  Vertical field runs are
  labeled by the prefix Z, zero net $z$ field runs are labeled by the
  prefix S, and vertical field runs are labeled by the prefix Y. 
The double brackets indicate that all
  data are time- and volume-averages at late times over the turbulent
  layer of the disk.  The magnetic field energy, Maxwell
  stress, and Reynolds stress are all normalized with respect to the
  initial gas pressure at the midplane, $P_0$. These runs include the
  effects of the Hall term and Ohmic dissipation.}
\end{deluxetable}

\begin{deluxetable}{lcccc}
\tablewidth{0.0pt}
\tablecolumns{5}
\tablecaption{Sano \& Stone  2002 - saturated pressure}
\tablehead{
  \colhead{Run} &
  \colhead{$\langle\langle P/P_0\rangle\rangle$} &
  \colhead{$\langle\langle P \rangle\rangle/\langle\langle B^2/8\pi\rangle\rangle$} &
  \colhead{$\langle\langle -B_xB_y\rangle\rangle/\langle\langle P\rangle\rangle\cdot $}&
  \colhead{$\langle\langle \rho v_x\delta v_y\rangle\rangle/\langle\langle P\rangle\rangle\cdot {2f(\Gamma)\over 3\Gamma}$}
}
\startdata 
Z02 & 2.7800E+00 & 7.8100E+00 & 1.3458E-02 & 2.8078E-03 \\
Z03 & 4.6300E+00 & 1.9500E+01 & 5.3869E-03 & 1.0874E-03 \\
Z04 & 2.6800E+00 & 3.2000E+01 & 3.3952E-03 & 1.0685E-03 \\
S01 & 2.5800E+00 & 5.5500E+01 & 1.5038E-03 & 1.1637E-03 \\
S03 & 1.9400E+00 & 1.5000E+01 & 6.1901E-03 & 1.5832E-03 \\
S10 & 3.3200E+00 & 3.9800E+01 & 2.2677E-03 & 6.1283E-04 \\
Y02 & 6.4400E+00 & 8.1700E+01 & 1.2659E-03 & 2.9872E-04 \\
Y04 & 5.8300E+00 & 4.6600E+01 & 2.1430E-03 & 4.9515E-04 \\
\enddata
\tablecomments{Runs with adiabatic equations of state.  Vertical field runs are
  labeled by the prefix Z, zero net $z$ field runs are labeled by the
  prefix S, and vertical field runs are labeled by the prefix Y. The double brackets indicate that all   data are time- and volume-averages at late times over the turbulent
  layer of the disk.  In this table, $P$ denotes the actual pressure at the
  time and position of averaging. These runs include the effects of
  the Hall term and Ohmic dissipation.}
\end{deluxetable}

\begin{deluxetable}{lccc}
\tablewidth{0.0pt}
\tablecolumns{4}
\tablecaption{Stone et al. 1996}
\tablehead{
  \colhead{Run} &
  \colhead{$\langle\langle 8 \pi P_0/B^2\rangle\rangle$} &
  \colhead{$(2 f(\Gamma)/3\Gamma)\langle\langle {- B_x B_y \over  4 \pi P_0}\rangle\rangle$} &  
  \colhead{$(2 f(\Gamma)/3\Gamma)\langle\langle\rho v_x \delta v_y / P_0\rangle\rangle$}}
\startdata 
 AZ1 & 42.2 & 0.001769001 & 0.000397217 \\
 AZ6 & 27.2 & 0.00327935  & 0.000739008 \\
 AY1 & 21.8 & 0.002817469 & 0.000672036 \\
 AL6 & 58.5 & 0.001129297 & 0.000274819 \\
\enddata
\tablecomments{Runs with adiabatic equations of state.  Zero net $z$ field runs
  are labeled by the prefix Z.  AY1 is a vertical field run and AL6
  has a ``flux loops'' initial field configuration. The double brackets indicate that all data are time-
  and volume-averages at late times over the turbulent layer of the
  disk.  The magnetic field energy, Maxwell stress, and
  Reynolds stress are all normalized with respect to the initial gas
  pressure at the midplane, $P_0$.}
\end{deluxetable}


\begin{deluxetable}{lccc}
\tablewidth{0.0pt}
\tablecolumns{4}
\tablecaption{Fleming \& Stone 2003}
\tablehead{
  \colhead{Run} &
  \colhead{$\langle\langle 8 \pi P(0)/B^2\rangle\rangle$} &
  \colhead{$(2 f(\Gamma)/3\Gamma)\langle\langle - B_x B_y / 4 \pi P(0)\rangle\rangle$} &  
  \colhead{$(2 f(\Gamma)/3\Gamma)\langle\langle\rho v_x \delta v_y / P(0)\rangle\rangle$}}

\startdata 
 Z1 & 6.6667E+01 & 2.3570E-03 & 4.7140E-04 \\
 Z2 & 3.3333E+02 & 4.7140E-04 & 1.4142E-04 \\
 Z3 & 1.0000E+03 & 2.8284E-04 & 4.7140E-05 \\
 Z4 & 3.5714E+01 & 4.2426E-03 & 1.4142E-03 \\
 Y1 & 5.0000E+02 & 3.2998E-04 & 4.7140E-05 \\
 Y2 & 8.3333E+01 & 1.8856E-03 & 4.7140E-04 \\
\enddata
\tablecomments{Runs with isothermal equations of state.  Vertical field
  runs are labeled by the prefix Z and toroidal field
  runs are labeled by the prefix Y. The double brackets indicate that all data are time- and
  volume-averages at late times over the turbulent layer of the disk:
  $0.4 < |z/H|<2$.  (These models also include a central ``dead'' zone
  ($0 < |z/H| < 0.4$) which is not included in the averaging.)  The
  magnetic field energy, Maxwell stress, and Reynolds stress are all
  normalized with respect to the midplane gas pressure, $P(0)$.  In
  these models, the ionization fraction varies with height.}
\end{deluxetable}

\begin{deluxetable}{lccc}
\tablewidth{0.0pt}
\tablecolumns{4}
\tablecaption{Miller \& Stone 2000}
\tablehead{
  \colhead{Run} &
  \colhead{$\langle\langle 8 \pi P(0)/B^2\rangle\rangle$} &
  \colhead{$(2 f(\Gamma)/3\Gamma)\langle\langle - B_x B_y / 4 \pi P(0)\rangle\rangle$} &  
  \colhead{$(2 f(\Gamma)/3\Gamma)\langle\langle\rho v_x \delta v_y / P(0)\rangle\rangle$}
}
\startdata 
 BY1 & 1.1547E+01 & 1.1125E-02 & 2.7483E-03 \\
 BY2 & 1.5773E+01 & 8.3910E-03 & 2.0553E-03 \\
 BY3 & 3.8314E+01 & 3.3564E-03 & 1.0324E-03 \\
 BY4 & 1.7036E+01 & 7.2596E-03 & 2.0600E-03 \\
 BY5 & 1.8832E+01 & 6.4582E-03 & 1.6546E-03 \\
 BY6 & 1.0627E+01 & 1.1597E-02 & 2.8426E-03 \\
 BYR1 & 1.7575E+01 & 6.7411E-03 & 1.8243E-03 \\
 BYR2 & 1.8018E+01 & 5.2326E-03 & 1.3435E-03 \\
 BYR3 & 9.1743E+00 & 1.6641E-02 & 4.0447E-03 \\
 ZN1 & 4.5455E+01 & 2.3099E-04 & 1.0842E-04 \\
 ZN2 & 2.6882E+02 & 2.5126E-04 & 1.2681E-04 \\
 BZ1 & 5.5804E-01 & 4.2756E-01 & 2.6823E-02 \\
 BZ2 & 1.1351E+00 & 1.4519E-01 & 1.6735E-02 \\
 BZ3 & 9.6712E-01 & 1.3114E-01 & 1.4802E-02 \\
\enddata
\tablecomments{Runs with isothermal equations of state.  Toroidal field runs are
  labeled by the prefix BY, zero net $z$ field runs are labeled by the
  prefix ZN, and pure $z$ field runs are labeled by the prefix BZ.
  ``R'' identifies the resistive runs.  The double brackets indicate that all data are time- and
  volume-averages at late times over the turbulent layer of the disk:
  $0 < |z/H| < 2$.  (These models also include an extended, quiescent
  corona ($2<|Z/H|<4$) which is not included in the averaging.)  The
  magnetic field energy, Maxwell stress, and Reynolds stress are all
  normalized with respect to the midplane gas pressure, $P(0)$.  These
  runs include the effects of Maxwell's displacement current.}
\end{deluxetable}

\begin{deluxetable}{lccc}
\tablewidth{0.0pt}
\tablecolumns{4}
\tablecaption{Stone et al. 1996}
\tablehead{
  \colhead{Run} &
  \colhead{$\langle\langle 8 \pi P(0)/B^2\rangle\rangle$} &
  \colhead{$(2 f(\Gamma)/3\Gamma)\langle\langle - B_x B_y / 4 \pi P(0)\rangle\rangle$} &  
  \colhead{$(2 f(\Gamma)/3\Gamma)\langle\langle\rho v_x \delta v_y / P(0)\rangle\rangle$}
}
\startdata 
 IZ1 & 6.54E+01 & 2.0930E-03 & 5.8926E-04 \\
 IZ2 & 2.50E+02 & 4.5726E-04 & 2.1213E-04 \\
 IZ3 & 6.29E+01 & 2.0836E-03 & 5.9397E-04 \\
 IZ6 & 4.98E+01 & 3.2150E-03 & 8.9567E-04 \\
 IY1 & 2.79E+01 & 4.6716E-03 & 1.1314E-03 \\
 IY2 & 1.63E+02 & 4.4312E-04 & 1.4991E-04 \\
 IY3 & 2.58E+01 & 3.7618E-03 & 9.7109E-04 \\
\enddata
\tablecomments{Runs with isothermal equations of state.  Toroidal field runs are
  labeled by the prefix IY and zero net $z$ field runs are labeled by
  the prefix IZ.  The double brackets indicate that all data are time- and volume-averages over the disk
  at late times.  The magnetic field energy, Maxwell stress, and
  Reynolds stress are all normalized with respect to the midplane gas
  pressure, $P(0)$.}
\end{deluxetable}

\eject

\begin{figure}
\centerline{
\epsfig{file=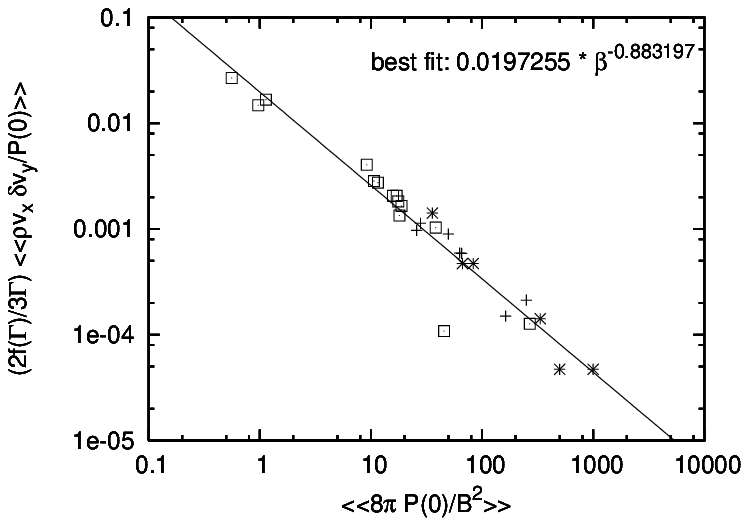,width=9.5cm}
\epsfig{file=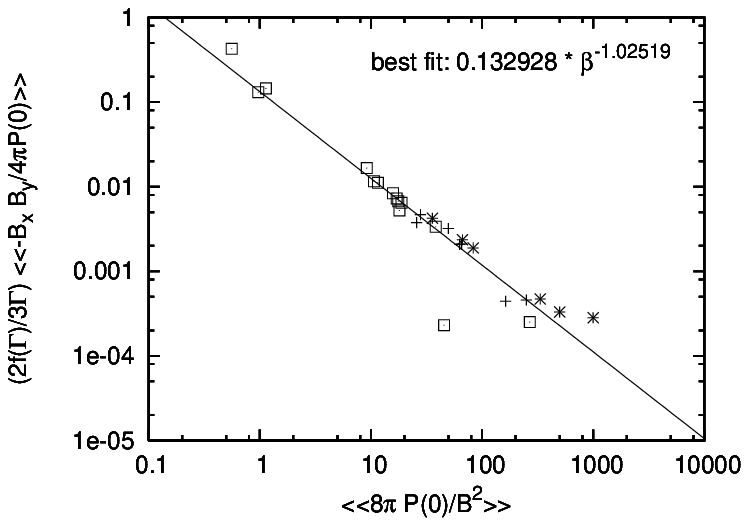,width=9.5cm}}
\centerline{
\epsfig{file=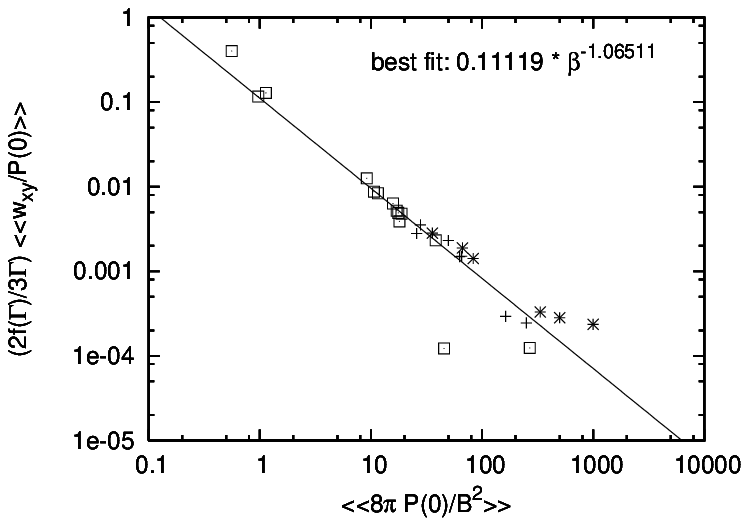,width=9.5cm}
\epsfig{file=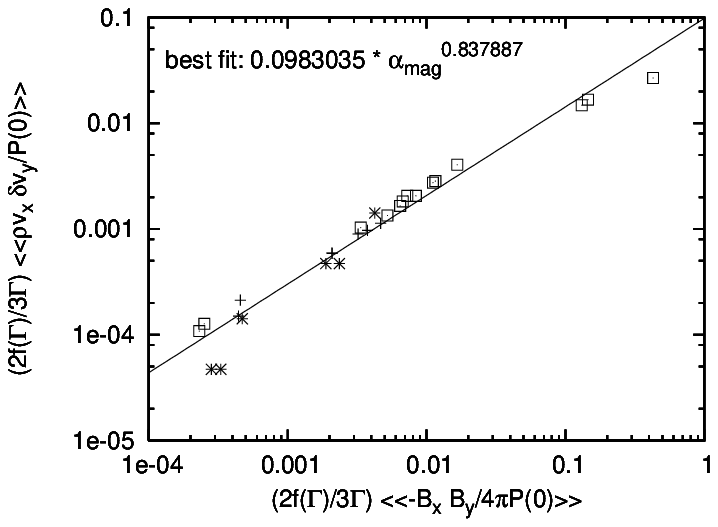,width=9.5cm}}
\caption{
Isothermal data of Tables 8-10. 
and best fit lines.
Top row:  $Log \ \alpha_{kin}(P(0))$ and $Log\ \alpha_{mag}(P(0))$ 
vs. $Log\ \beta(P(0))$ respectively.
Bottom row: $Log \ \alpha_{tot}(P(0))$ vs. $Log \ \beta(P(0))$
and $Log\ \alpha_{kin}(P(0))$
vs. $Log\ \alpha_{mag}(P(0))$.
The double brackets indicate that the data represent a combination of 
 spatial average and  late time average (e.g. after 15 orbits). 
The symbols indicate the following specific data sets respectively: 
$*=$ Fleming and Stone (2003);  
$\boxdot=$ Miller and Stone (2000);  
$+=$ Stone et al. (1996).  
Values of the magnetic
  field energy, Maxwell stress, and Reynolds stress are all normalized
  with respect to the midplane gas pressure, $P(0)$, and for all of
  these runs, $P(0)=5\cdot 10^{-7}$.
For the most part, 
despite the different initial
 and boundary conditions and wide
ranges of $\alpha_{mag}$, $\alpha_{kin}$, and $\alpha_{tot}$,
products of  form $\alpha\beta$  lie close to the best fit lines shown.
The line equations are at the top of each panel.}
\label{fig:abkinmag}
\end{figure}

\begin{figure}
\centerline{
\epsfig{file=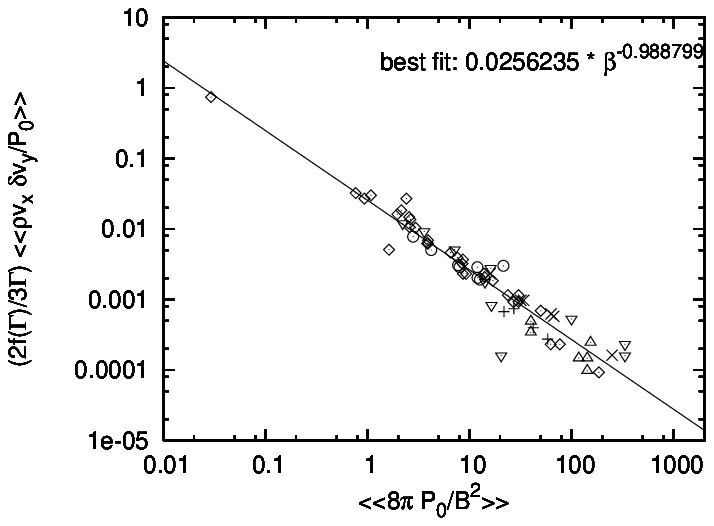,width=9.5cm}
\epsfig{file=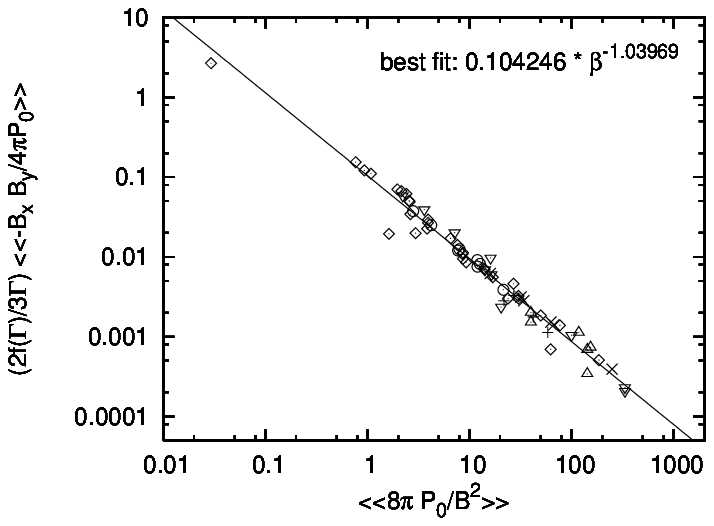,width=9.5cm}}
\centerline{
\epsfig{file=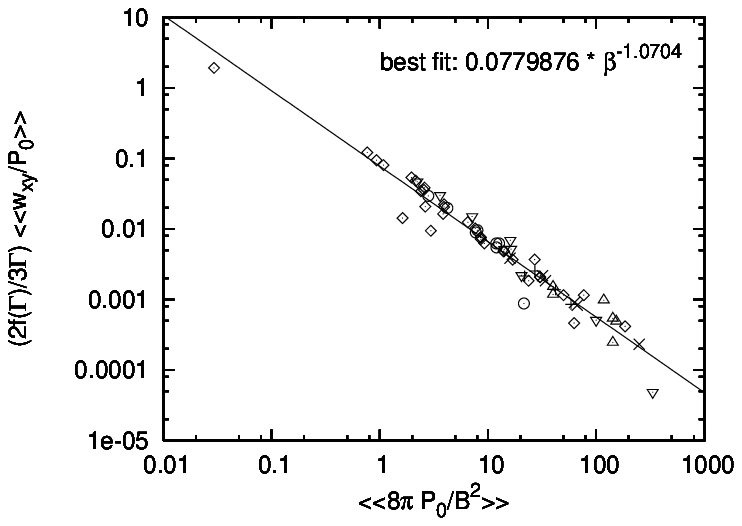,width=9.5cm}
\epsfig{file=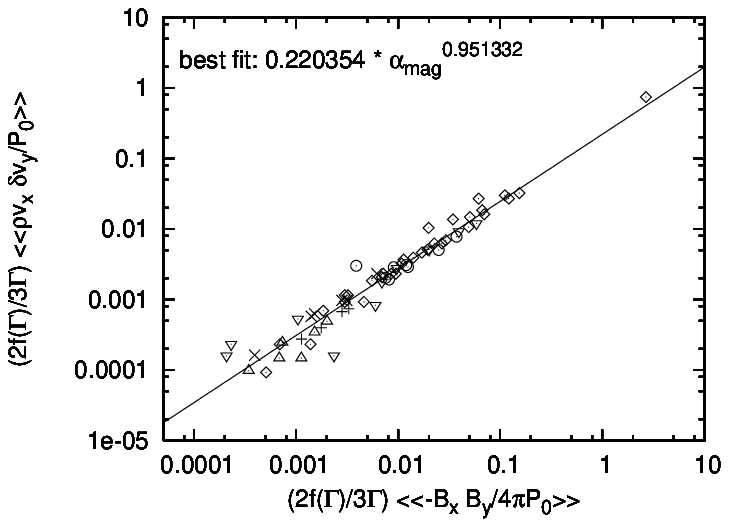,width=9.5cm}}
\caption{
Adiabatic runs from  Tables 1-5,7 and the best-fit lines.  
Top row:  $Log \ \alpha_{kin}(P_0)$ and $Log\ \alpha_{mag}(P_0)$ 
vs $Log\ \beta(P_0)$ respectively.
Bottom row: $Log \ \alpha_{tot}(P_0)$ and $Log\ \alpha_{kin}(P_0)$ vs.
$Log\ \alpha_{mag}(P_0)$
The double bracket averages are as in Fig. 1.
 $P_0$ is the initial gas pressure at the midplane,
which differs from the final midplane pressure $P(0)$ for the adiabatic
case (unlike the isothermal case of Fig 1. where the two pressures are interchangable.). The symbols indicate the following data sets
respectively: 
$\triangle=$ Brandenburg et al. (1995); 
$\triangledown=$ Fleming, Stone, and Hawley (2000);  
$\lozenge=$ Hawley, Gammie, and Balbus (1995);
$\times=$ Hawley, Gammie, and Balbus (1996);
$\odot=$ Sano and Stone (2002) (includes a Hall term);
$+=$ Stone et al. (1996).
Equations for the best fit lines are given.}
\label{fig:abkinmag}
\end{figure} 

\eject



\begin{figure}
\centerline{
\epsfig{file=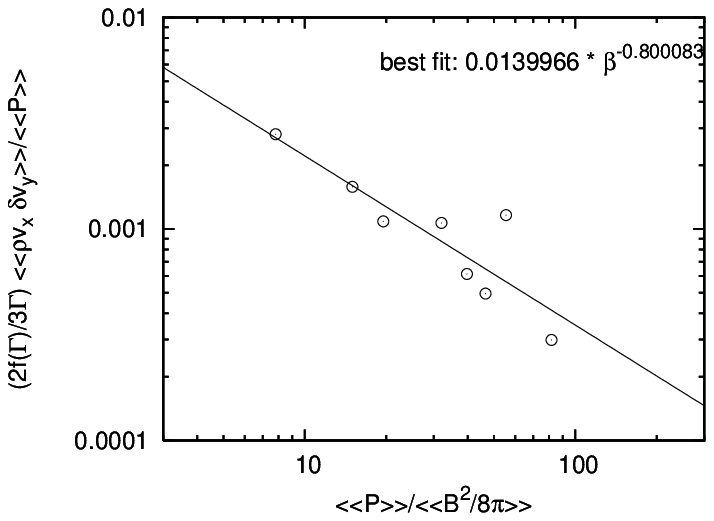,width=9.5cm}
\epsfig{file=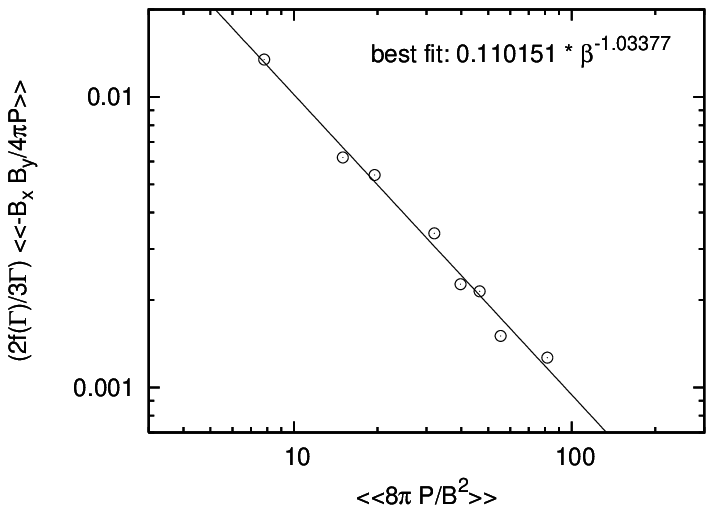,width=9.5cm}}
\centerline{
\epsfig{file=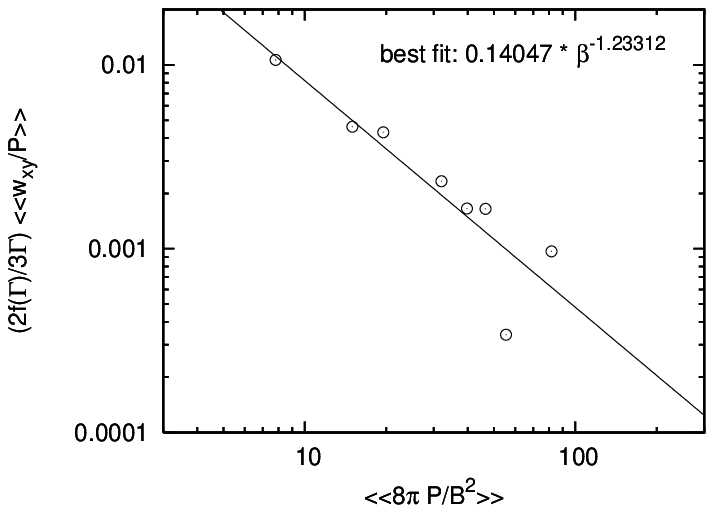,width=9.5cm}
\epsfig{file=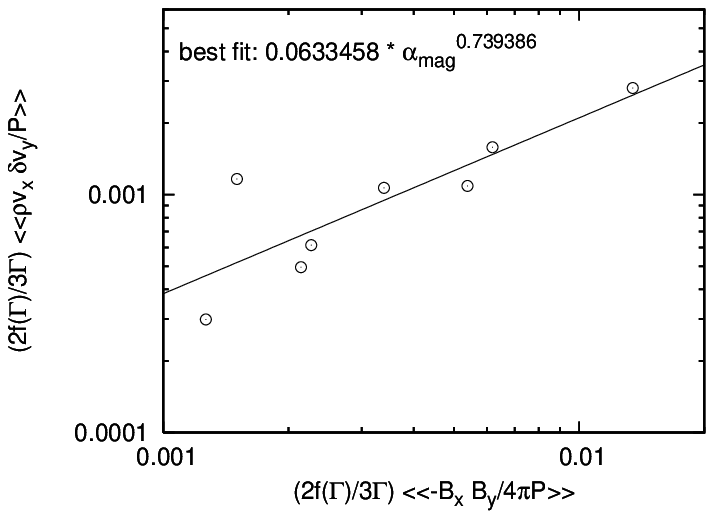,width=9.5cm}
}
\caption{
Same quantities plotted
as in Fig. 2 but for the available adiabatic runs  in which the 
saturated midplane pressure $P(0)$ is used rather than the initial midplane
pressure  $P_0$ of the previous figure. 
All data here are from Sano \& Stone (2002).}
\end{figure}

\eject


\end{document}